\documentclass[prl,superscriptaddress,twocolumn,%
  showpacs,preprintnumbers,amsmath,amssymb]{revtex4-1}
\usepackage[utf8]{inputenc}
\usepackage{graphicx}
\usepackage[breaklinks, colorlinks=true, citecolor=blue, linkcolor=cyan]{hyperref}
\usepackage{bm}
\usepackage{physics}

\newcommand{\p}{\partial}
\newcommand{\vOmega}{\mathfrak{w}}
\newcommand{\bs}{{\bm s}}
\newcommand{\bv}{{\bm v}}

\begin{document}

\title{Relativistic spin hydrodynamics with antisymmetric spin tensors\\ and an extension of the Bargmann-Michel-Telegdi equation}

\author{Shuo Fang}
\email{fangshuo@mail.ustc.edu.cn}
\affiliation{Department of Modern Physics, University of Science and Technology of China, Anhui 230026, China}
\affiliation{Technical University of Munich, TUM School of Natural Sciences, Physics Department, James-Franck-Str. 1, 85748 Garching, Germany}

\author{Kenji Fukushima}
\email{fuku@nt.phys.s.u-tokyo.ac.jp}
\affiliation{Department of Physics, The University of Tokyo, 7-3-1 Hongo, Bunkyo-ku, Tokyo 113-0033, Japan}

\author{Shi Pu}
\email{shipu@ustc.edu.cn}
\affiliation{Department of Modern Physics, University of Science and Technology of China, Anhui 230026, China}
\affiliation{Southern Center for Nuclear-Science Theory (SCNT), Institute of Modern Physics, Chinese Academy of Sciences, Huizhou 516000, Guangdong Province,
China}

\author{Dong-Lin Wang}
\email{donglinwang@mail.ustc.edu.cn}
\affiliation{Department of Modern Physics, University of Science and Technology of China, Anhui 230026, China}

\begin{abstract}
We derive a formulation of relativistic spin hydrodynamics with totally antisymmetric spin tensors that satisfy the Frenkel-Mathisson-Pirani condition.  In our proposed spin hydrodynamics, the second law of thermodynamics is fulfilled by the spin-induced corrections in the heat flow, the viscous tensor, and the antisymmetric part of the energy-momentum tensor.  These corrections are interpreted as the inverse spin Hall effect and the anomalous Hall effect in the nonrelativistic limit.  We show that our evolution equation for the spin density is interpreted as an extension of the Bargmann-Michel-Telegdi equation known in relativistic many-body systems, including the Thomas precession term, the spin-rotation term, and new coupling terms between spin and hydrodynamic variables.
\end{abstract}
\maketitle

\paragraph*{Introduction:}

Spin, a fundamental building block of modern physics, was discovered nearly a century ago and has held a crucial position in both theoretical and experimental physics.  In high-energy physics, several pivotal measurements are intricately tied to the concept of spin, including the proton spin crisis (see reviews~\cite{Leader:2013jra,Deur:2018roz,Ji:2020ena}), single spin asymmetry and transverse polarization of hyperons (see reviews~\cite{DAlesio:2007bjf,Aidala:2012mv}), the neutron electric dipole moment~\cite{Abel:2020pzs}, the muon anomalous magnetic dipole moment~\cite{Muong-2:2021ojo} and others.
Recently, the study of spin physics has expanded into relativistic many-body systems. In relativistic heavy-ion collisions, a variety of novel spin-related phenomena have been extensively investigated, including interactions between spin and electromagnetic fields, particularly the chiral magnetic effect and the chiral separation effect~\cite{Kharzeev:2007jp,Fukushima:2008xe,Kharzeev:2015znc}, as well as the conversion between spin and orbital angular momentum, as evidenced by spin polarization~\cite{Liang:2004ph,STAR:2017ckg} and spin alignment~\cite{Liang:2004xn,STAR:2022fan}.  Understanding spin dynamics in hot and dense matter is crucial for analyzing experimental data from relativistic heavy-ion collisions.

Although the spin dynamics of individual particles has been extensively investigated with the Dirac equation~\cite{Leader_2001}, the relativistic characterization of spin evolution in many-body quantum systems remains only partially established.  This is because of the relativistic nature of spin, which is no longer conserved, unlike in nonrelativistic systems.
Thus, it is a nontrivial question how to find the evolution equation of the macroscopic spin degrees of freedom in many-body systems.

Relativistic spin hydrodynamics has been motivated by experiments to probe many-body quantum systems with relativistic heavy-ion collisions.  There, the initial orbital angular momentum of $\sim 10^5\hbar$ stems from the noncentral collision of two nuclei~\cite{Jiang:2016woz}, and has been transferred to the spin polarization of quarks through the spin-orbit coupling~\cite{Liang:2004ph}.  Polarization effects ultimately result in measurable observables such as spin polarization of $\Lambda$ hyperons~\cite{STAR:2017ckg,STAR:2019erd} and spin alignment of vector mesons~\cite{STAR:2022fan,ALICE:2022dyy}.
The dynamical evolution of spin is essential to understand the experimental data.  Although some experimental results, e.g.\ the global polarization of $\Lambda$ hyperons induced by thermal vorticity, have been successfully explained by theory~\cite{Becattini:2013fla}, challenges remain in fully elucidating phenomena such as the local polarization.

Steady progress has been made in relativistic spin hydrodynamics since the early stage of formulation~\cite{Becattini:2009wh,Florkowski:2017ruc}; see studies based on the entropy principle~\cite{Hattori:2019lfp,Fukushima:2020ucl,Li:2020eon}, the kinetic theory approach~\cite{Florkowski:2018fap,Weickgenannt:2022zxs}, and the effective theories~\cite{Montenegro:2017rbu}.  Yet, a comprehensive understanding of spin hydrodynamics remains elusive.  Let us point out that two vital features of spin dynamics are often dismissed in these approaches.  First, consistency with the covariant spin equation of motion in a classical fluid has received only limited attention so far.  Second, a connection between relativistic spin hydrodynamics and well-known spin phenomena in nonrelativistic systems, such as  the inverse spin Hall effect~\cite{sinova2015spin} and the anomalous Hall effect~\cite{Xiao:2009rm,nagaosa2010anomalous}, has not been established.  These two features hold significant importance in extending the applicability of relativistic spin hydrodynamics from heavy-ion collisions to broader scientific domains.

Regarding spin motion, under the magnetic field that plays a role similar to rotation, the precession and relaxation of classical spin can be well described by the Landau-Lifshitz-Gilbert equation~\cite{lakshmanan2011fascinating}.  The precession equation has been generalized to a covariant form, that is, the Bargmann-Michel-Telegdi (BMT) equation~\cite{Jackson:1998nia}.  Using a fluid velocity vector, $u^\mu$, one can write the BMT equation as
\begin{equation}
    \dot{s}^\mu = -u^\mu s^\nu \dot{u}_\nu
    + \gamma \Delta^{\mu\rho} F_{\rho\nu} s^\nu \,,
    \label{eq:BMT}
\end{equation}
where the first term represents the Thomas precession, and the second term $\propto\gamma$ (where $\gamma$ is the gyromagnetic ratio) causes the magnetic-spin precession.  We will discuss that the spin-rotation coupling should give rise to similar terms.  In the above expression, the time derivative is $\dot{A}:=u_\mu \partial^\mu A$ and the spatial projection is $\Delta^{\mu\nu}:=g^{\mu\nu}-u^\mu u^\nu$.  It is not clear how to prescribe a systematic generalization of the BMT equation with further coupling terms between spin, $s^\mu$, and other hydrodynamic variables.  In the present work, we will demonstrate that our new formulation of relativistic spin hydrodynamics incorporates a natural extension of the BMT equation within the framework of viscous hydrodynamics, as well as the inverse spin Hall effect and the anomalous Hall effect.

In this study, we undertake the derivation of relativistic spin hydrodynamics based on the entropy principle, taking into account totally antisymmetric spin tensors.  In our convention, the metric is $g_{\mu\nu}=\mathrm{diag}(+,-,-,-)$ and the Levi-Civita symbol is $\varepsilon^{0123}=-\varepsilon_{0123}=+1$.  For a generic tensor $A^{\mu\nu}$, we also use the following standard notations:
$A^{(\mu\nu)}:=(A^{\mu\nu}+A^{\nu\mu})/2$, 
$A^{[\mu\nu]}:=(A^{\mu\nu}-A^{\nu\mu})/2$,
and $A^{<\mu\nu>}:=\Delta^{\alpha(\mu}\Delta^{\nu)\beta}A_{\alpha\beta}-\frac{1}{3}\Delta^{\mu\nu}(\Delta^{\alpha\beta}A_{\alpha\beta})$. 
\vspace{0.5em}

\paragraph*{Spin tensor and spin density:}

The rotational symmetry for spin-$1/2$ fields is associated with the conservation law for the expectation value of the total canonical angular momentum operator,
\begin{align}
  J^{\lambda\mu\nu} = L^{\lambda\mu\nu}+\Sigma^{\lambda\mu\nu}\,,
  \label{eq:J_TAM}
\end{align}
where $L^{\lambda\mu\nu}=x^{\mu}\Theta^{\lambda\nu}-x^{\nu}\Theta^{\lambda\mu}$ is the canonical orbital angular momentum with $\Theta^{\mu\nu}$ the canonical energy-momentum tensor and $\Sigma^{\lambda\mu\nu}$ is the canonical spin tensor.

The physical meaning of $\Sigma^{\lambda\mu\nu}$ can be understood in the particle rest frame where the components, $\varepsilon^{ijk}\Sigma^{0jk}$, are identified as the conventional spin three-vector defined in quantum mechanics.  We note that Lorentz group has no unitary representation unless the dimension of the representation is infinite, that is the case for the one generated by the Hermitian generators of orbital angular momenta.
Thus, the finite-dimensional spin-boost, $\Sigma^{00i}$ in the rest frame or $u_\mu u_\nu \Sigma^{\mu\nu i}$ in general is not Hermitian and thus not a physical observable.  There are several well-known prescriptions to eliminate redundant components in $\Sigma^{\lambda\mu\nu}$.
In this work, we employ the following strategy to utilize an antisymmetrized form of the Dirac Lagrangian leading to a Hermitian spin tensor, $\Sigma^{\lambda\mu\nu}=\frac{i}{8}\bar{\psi}\{\gamma^{\lambda},[\gamma^{\mu},\gamma^{\nu}]\}\psi$.  Interestingly,  this form of spin tensor is totally antisymmetric with respect to $\lambda$, $\mu$, $\nu$ which is immediately confirmed from $\{\gamma^\lambda,[\gamma^\mu,\gamma^\nu]\}=\{ [\gamma^\lambda,\gamma^\mu],\gamma^\nu\}=\{\gamma^\mu,[\gamma^\nu,\gamma^\lambda]\}$.
Obviously, $\Sigma^{00i}=0$ from the symmetry property.

We comment that in the preceding literature~\cite{Hattori:2019lfp,Fukushima:2020ucl} another spin tensor was introduced. Although $\Sigma^{\lambda\mu\nu}$ as defined above and another form are equivalent through the equation of motion, neither the spin nor the total angular momentum in the previous works possesses the Hermitian property.  Furthermore, recent research has revealed that constructing a framework of consistent and stable spin hydrodynamics along these lines~\cite{Hattori:2019lfp,Fukushima:2020ucl} poses difficulties; see Ref.~\cite{Xie:2023gbo} and references therein.
\vspace{0.5em}

\paragraph*{Conservation equations:}

For notational brevity, henceforth, we shall refer to the expectation values of the spin tensor and the total angular momentum operators as $\Sigma^{\lambda\mu\nu}$ and $J^{\lambda\mu\nu}$ without the brackets.

The main conservation equations for spin hydrodynamics are written for the energy-momentum $\Theta^{\mu\nu}$, charge current $j^{\mu}$, and total angular momentum $J^{\lambda\mu\nu}$ as
\begin{equation}
  \partial_{\mu}\Theta^{\mu\nu}=0\,,
  \qquad
  \partial_{\mu}j^{\mu}=0\,,
  \qquad
  \partial_{\lambda}J^{\lambda\mu\nu}=0\,.
  \label{eq:ConservationEqs}
\end{equation}
The constitutive equations for $\Theta^{\mu\nu}$ and $j^{\mu}$ are expressed with irreducible variables in the following tensor decomposition:
\begin{align}
  \Theta^{\mu\nu} = \Theta^{(\mu\nu)}+2q^{[\mu}u^{\nu]}+\phi^{\mu\nu}\,,
  \qquad
  j^{\mu}=nu^{\mu}+\nu^{\mu}\,,
  \label{eq:constittuive_eq}
\end{align}
where $\Theta^{(\mu\nu)}=(e+p)u^{\mu}u^{\nu}-pg^{\mu\nu}+2h^{(\mu}u^{\nu)}+\pi^{\mu\nu}$ stands for the symmetric part of $\Theta^{\mu\nu}$.  Here, we introduce standard variables in relativistic hydrodynamics such as the heat flow $h^\mu$, the viscous tensor $\pi^{\mu\nu}$, the velocity of fluid cells $u^\mu$, and the particle diffusion current $\nu^\mu$.  We note that $u\cdot q=(u\cdot \phi)^\mu=u\cdot h=(u\cdot\pi)^\mu=0$, so that the tensor decomposition is unique.  Thermodynamic quantities are the energy density $e$, the particle number density $n$, and the pressure $p$.
The conservation laws of the energy-momentum tensor and the angular momentum are coupled to conclude
\begin{equation}
  \partial_{\lambda}\Sigma^{\lambda\mu\nu}
  = -2\Theta^{[\mu\nu]}
  = -2 (2q^{[\mu} u^{\nu]} + \phi^{\mu\nu})\,,
  \label{eq:SpinEvEq}
\end{equation}
indicating that the antisymmetric part of the energy-momentum tensors violates the spin conservation.

To include spin in hydrodynamics, we introduce the spin potential $\omega_{\mu\nu}$ and modify the thermodynamic relations as follows:
\begin{equation}
  e+p = \beta^{-1} s + \mu n + \omega_{\mu\nu}S^{\mu\nu}\,,
  \label{eq:ThermoRela1}
\end{equation}
where $\beta=1/T$ is the inverse temperature and $s$ is the entropy density.  The Gibbs relations read $de=\beta^{-1}ds+\mu dn+\omega_{\mu\nu}dS^{\mu\nu}$ and $dp=s d\beta^{-1}+nd\mu+S^{\mu\nu}d\omega_{\mu\nu}$.  In this work, we assume that the spin effects are significant and follow the power counting scheme in our previous work~\cite{Fukushima:2020ucl}, i.e., $S^{\mu\nu}\sim \mathcal{O}(1)$ and $\omega_{\mu\nu}\sim\mathcal{O}(\hbar)\sim \mathcal{O}(\partial)$.
We emphasize that this counting can be consistent with thermodynamics;  a correction to $p$ by $S^{\mu\nu}\omega_{\mu\nu}$ can be treated as $\mathcal{O}(\hbar)$.
At global equilibrium, the Killing vector condition \citep{Becattini:2012tc} leads to
\begin{equation}
    \frac{1}{2\beta} \Omega^{\mu\nu} \biggr|_{\text{Global Equil.}} = \omega^{\mu\nu}\,,
    \label{eq:GE}
\end{equation}
where $\Omega^{\mu\nu} := -\partial^{[\mu} (\beta u^{\nu]} )$ is the thermal vorticity tensor.  This relation is given an interpretation that the fluid motion follows the local vorticity at equilibrium.
\vspace{0.5em}

\paragraph*{Entropy principle:}

Let us quickly review the main results in the previous works~\cite{Hattori:2019lfp,Fukushima:2020ucl}.  Previously, a factorized form of spin, $\Sigma^{\lambda\mu\nu}=u^\lambda S^{\mu\nu}$, was adopted.  Then, the divergence of the entropy density is expressed in a simple form, and from the condition of non-negative entropy generation, it was found that $q^\mu \propto (\Omega^{\mu\nu}-2\beta\omega^{\mu\nu})u_\nu$ and $\phi^{\mu\nu}\propto \Delta^{\mu\alpha}\Delta^{\nu\beta} (\Omega_{\alpha\beta} - 2\beta\omega_{\alpha\beta})$ with the proportionality constants called the boost heat conductivity and the rotational viscosity, respectively.  Clearly, these dissipative terms disappear once the global equilibrium condition~\eqref{eq:GE} is fulfilled.

Here, we shall introduce the local spin density as $S^{\mu\nu} = u_\lambda \Sigma^{\lambda\mu\nu}$ using totally antisymmetric $\Sigma^{\lambda\mu\nu}$.  For some former attempts to implement antisymmetric $\Sigma^{\lambda\mu\nu}$, see Refs.~\cite{Hongo:2021ona,Cao:2022aku}.  It is important to note that $(u\cdot S)^\mu=0$ is derived from $u_\lambda u_\mu \Sigma^{\lambda\mu\nu}=0$.  This relation is called the Frenkel-Mathisson-Pirani condition, which is consistent with the treatment of spin in terms of the normalized Pauli-Luba\'{n}ski vector.  In the particle motion in general relativity, other prescriptions such as the Tulczyjew-Dixon condition, the Newton-Wigner condition, the Ohashi-Kyrian-Semer\'ak condition are known depending on the choice of the representative worldline~\cite{Costa:2017kdr}.  In hydrodynamics, however, only the Frenkel-Mathisson-Pirani condition is possible in terms of hydrodynamic variables.

Now, we can antisymmetrize the factorized form and decompose the spin tensors as $\Sigma^{\lambda\mu\nu}  =  u^{\lambda}S^{\mu\nu}+u^{\mu}S^{\nu\lambda}+u^{\nu}S^{\lambda\mu}+\Sigma_{(1)}^{\lambda\mu\nu}$, where the last term $\Sigma_{(1)}^{\lambda\mu\nu}$ is a totally antisymmetric tensor of $\mathcal{O}(\partial)$ satisfying $u_\lambda \Sigma_{(1)}^{\lambda\mu\nu}=0$.  It is a straightforward exercise to find the following relation,
\begin{equation}
  q^\mu
  = -\frac{1}{2} u_\nu \partial_\lambda \Sigma^{\lambda\mu\nu}
  = \frac{1}{2} \bigl( S^{\mu\nu} \dot{u}_\nu + \Delta^{\mu}_{\nu} \partial_\lambda S^{\nu\lambda} \bigr)\,,
  \label{eq:qmu}
\end{equation}
using $u_\nu\partial_\lambda u^\nu=0$ and $u_\mu\partial_\lambda S^{\mu\nu}=-S^{\mu\nu} \partial_\lambda u_\mu$.

We are ready to consider the entropy production.  As postulated previously~\citep{Fukushima:2020ucl}, the generalized entropy current is $\mathcal{S}^{\mu}=\beta ( u_{\nu}\Theta^{\mu\nu}+pu^{\mu}-\mu j^{\mu}-\omega_{\rho\sigma}S^{\rho\sigma}u^{\mu} ) + \mathcal{O}(\partial^2)$.  The divergence of the entropy current simplifies as
\begin{align}
  \partial_\mu \mathcal{S}^\mu
  &= \bigl[ 2h^{(\mu} u^{\nu)} + \pi^{\mu\nu} + 2q^{[\mu} u^{\nu]}
    + \phi^{\mu\nu} \bigr]\partial_\mu (\beta u_\nu) \notag\\
  &\qquad - \nu^\mu \partial_\mu (\beta\mu)
    -\partial_\mu (u^\mu S^{\rho\sigma}) \beta\omega_{\rho\sigma} \notag\\
  &= (h^\mu \!-\! \mathcal{H}\nu^\mu) (\partial_\mu\beta \!+\! \beta \dot{u}_\mu)
    + \beta \pi^{\mu\nu}
    \partial_{(\mu} u_{\nu)}\notag\\
  &\qquad + 2\beta \omega_{\mu\nu} S^{\lambda\mu} \partial_\lambda u^\nu
  + q^\mu (\partial_\mu\beta - \beta \dot{u}_\mu)
     \notag\\ 
  &\qquad + \phi^{\mu\nu} (2\beta\omega_{\mu\nu} + \partial_{[\mu} \beta u_{\nu]} )
    + \mathcal{O}(\partial^3) \,,
    \label{eq:divS}
\end{align}
where $\mathcal{H}=(e+p)/n$ is the enthalpy per particle and we used $(u\cdot \omega)^\mu=0$, which is a natural requisite since our new $S^{\mu\nu}$ satisfies $(u\cdot S)^\mu=0$.
We note that $\partial_{(\mu}u_{\nu)}$ in a term coupled with $\pi^{\mu\nu}$ can be further decomposed as $\partial_{<\mu} u_{\nu>} + (\Delta_{\mu\nu}/3)(\partial\cdot u)$.
In the above, the first line contains the conventional dissipative terms with $h^\mu$, $\nu^\mu$, and $\pi^{\mu\nu}$ in relativistic hydrodynamics.  The last line with $\phi^{\mu\nu}$ has been previously documented in Refs.~\cite{Hattori:2019lfp,Fukushima:2020ucl}.
The important difference lies in the second line with $\omega_{\mu\nu}$ and $q^\mu$.

Previously, $q^\mu$ was treated to be fixed from the second law of thermodynamics, i.e., $\partial_\mu \mathcal{S}^\mu \geq 0$.  In the present strategy, $q^\mu$ is already given in Eq.~\eqref{eq:qmu}.  Besides, there was no such a term of $2\beta\omega_{\mu\nu} S^{\lambda\mu}\partial_\lambda u^\nu$ because $\partial_\mu(u^\mu S^{\rho\sigma})$ was simply $\partial_\mu \Sigma^{\mu\rho\sigma}$ which was translated into $q^\mu$ and $\phi^{\mu\nu}$ terms via Eq.~\eqref{eq:SpinEvEq}.  Now, because of our altered definition of $\Sigma^{\lambda\mu\nu}$, this additional term emerges.  At a glance, it seems impossible to guarantee $\partial_\mu \mathcal{S}^\mu \geq 0$.  We will, however, demonstrate below that these unwanted terms with $\omega_{\mu\nu}$ and $q^\mu$ in the second line can be absorbed by \textit{renormalization} in other terms.

Our task of renormalization is to find a translation from Eq.~\eqref{eq:divS} to the following form:
\begin{align}
  &\partial_\mu (\mathcal{S}^\mu + \delta\mathcal{S}^\mu)
  = (h^\mu-\mathcal{H}\nu^\mu+h_s^\mu)(\partial_\mu\beta+\beta\dot{u}_{\mu}) \notag\\
  & + \beta(\pi^{\mu\nu}+\pi_{s}^{\mu\nu})\partial_{(\mu}u_{\nu)} \notag\\
  & + (\phi^{\mu\nu}+\phi_s^{\mu\nu})(2\beta\omega_{\mu\nu} + \partial_{[\mu} \beta u_{\nu]} )
    + \mathcal{O}(\partial^3)\,.
    \label{eq:divSren}
\end{align}
Then, the second law of thermodynamics is guaranteed by the modified constitutive equations parametrized as
\begin{subequations}
  \begin{align}
  h^\mu - \mathcal{H}\nu^\mu + h_s^\mu & = -\sigma\Delta^{\mu\nu}(\partial_{\nu}\beta + \beta \dot{u}_\nu)\,,
  \label{eq:dissipative_02_1}\\
  \pi^{\mu\nu} + \pi_s^{\mu\nu} & = \zeta\Delta^{\mu\nu}(\partial\cdot u) + \eta\partial^{<\mu}u^{\nu>}\,,
  \label{eq:dissipative_02_2}\\
  \phi^{\mu\nu} + \phi_s^{\mu\nu} & = \gamma_\phi\Delta^{\mu\rho}\Delta^{\nu\sigma}(2\beta\omega_{\rho\sigma} - \Omega_{\rho\sigma} )\,,
  \label{eq:dissipative_02}
  \end{align}
\end{subequations}
where $\sigma,\zeta,\eta$ are the non-negative transport coefficients corresponding to the conductivity, the bulk and shear viscosities, respectively, and $\gamma_\phi$ is another non-negative transport coefficient, i.e., the rotational viscosity.

The corrections, $h_s^\mu$, $\pi_s^{\mu\nu}$, and $\phi_s^{\mu\nu}$ depend on the prescription, and in the above example, we found:
\begin{subequations}
  \begin{align}
  h_s^\mu
  &= -\frac{1}{2} \Delta^\mu_\nu \partial_\lambda S^{\nu\lambda}
    + \frac{1}{2} S^{\mu\nu}\dot{u}_\nu \notag\\
  & \qquad + \xi S^{\mu\nu}(\beta^{-1}\partial_\nu\beta + \dot{u}_\nu)\,,\\
  \pi_s^{\mu\nu}
  &= -2S^{(\mu\lambda} \omega^{\nu)}_{\;\lambda}
    + 2 
    \Delta^{\mu\nu} S^{\rho\sigma}\omega_{\rho\sigma} v_n^2\,,\\
  \phi_s^{\mu\nu}
  &= -2S^{[\mu\lambda} \omega^{\nu]}_{\;\lambda}
    - \beta^{-1}\dot{\beta} S^{\mu\nu}\,.
    \label{eq:phi_s}
  \end{align}
\end{subequations}
Here, $\xi$ is a free parameter and $v_n^2:=(\p p/\p e)_{n,S^{\mu\nu}}$ that is the speed of sound squared at constant density.  For the derivation of these relations, see End Matter.  Below we shall discuss the physical meaning of $\phi_s^{\mu\nu}$ that does not involve $\xi$ and thus this part is a robust result in our analysis.  We also note that the last term in $\phi_s^{\mu\nu}$ will be replaced with $\beta^{-1}\dot{\beta}=v_n^2 (\partial\cdot u)$ (see End Matter) in later discussions but the above form is more convenient for the tensor decomposition.  Below, we shall closely discuss three implications.
\vspace{0.5em}

\noindent
\textbf{Modified BMT equations}
---
Multiplying the projectors $\Delta^{\mu\alpha}\Delta^{\nu\beta}$ to both sides of the conservation equation~\eqref{eq:SpinEvEq} and inserting Eqs.~\eqref{eq:dissipative_02} and \eqref{eq:phi_s}, we derive the evolution equation for spin density as
\begin{align}
  \dot{s}^\mu & = -u^\mu s^\nu \dot{u}_\nu
    + (\varepsilon^{\mu\nu\rho\sigma} s_\nu u_\rho - 2\beta\gamma_\phi g^{\mu\sigma}) (2\omega_\sigma - \vOmega_\sigma) \notag \\
   &\quad -s_\nu \partial^{<\mu}u^{\nu>} - \Bigl( \frac{1}{3} + 2v_n^2 \Bigr) s^\mu(\partial\cdot u) \,,
   \label{eq:MOdifiedBMT}
\end{align}
where we introduced the spin vector, $s^\mu$, satisfying $u\cdot s=0$, as
\begin{equation}
  s^\mu := -\frac{1}{2}\varepsilon^{\mu\nu\rho\sigma}
  u_\nu S_{\rho\sigma}\,,
\end{equation}
from which we immediately recover $S^{\mu\nu} = -\varepsilon^{\mu\nu\rho\sigma} u_\rho s_\sigma$.  Also, in the same way, we defined the spin potential vector and the vorticity vector, respectively, as
\begin{equation}
    \omega^{\mu} := -\frac{1}{2}  \varepsilon^{\mu\nu\rho\sigma} u_\nu \omega_{\rho\sigma}\,,\quad
    \vOmega^{\mu} := \frac{1}{2}\varepsilon^{\mu\nu\rho\sigma}u_{\nu}\partial_{\rho}u_{\sigma}\,,
\end{equation}
which reduces to the classical
vorticity vector $\vOmega^{\mu}\rightarrow(0,\,\frac{1}{2}\nabla\times\bv)$
in the fluid rest frame.

Equation~\eqref{eq:MOdifiedBMT} can be construed as an adaptation of the BMT equation in the context
of many-body systems. The first term $-u^\mu s^\nu \dot{u}_\nu$ corresponds to the Thomas precession \cite{Weinberg:1972kfs}, which appears consistently for antisymmetrized spin tensors.  In the second term, $\varepsilon^{\mu\nu\rho\sigma}s_\nu u_\rho \omega_\sigma \rightarrow - {\bm \omega} \times {\bm s}$ is the spin-rotation coupling term, which is a rotation counterpart of the spin-orbit coupling, $-{\bm B}\times {\bm s}$, in the BMT equation~\eqref{eq:BMT}.  This term should naturally arise from the rotation-induced Hamiltonian, $H_\omega = - {\bm \omega} \cdot {\bm s}$, and this term is already incorporated in the thermodynamic relation~\eqref{eq:ThermoRela1}.  We emphasize that the spin motion must have such a term derived from the Heisenberg equation, and in our formulation $\phi_s^{\mu\nu}$ generates this coupling.  Therefore, the presence of this rotation-induced precession term is a justification for our formulation.
The other terms are new corrections from relativistic spin hydrodynamics.  We observe that the rotation-induced precession disappears at global equilibrium due to the combination of $(\vOmega_\sigma - 2\omega_\sigma)$ which is vanishing when Eq.~\eqref{eq:GE} is achieved.  It is also interesting that the spin along $(2\omega_\sigma - \vOmega_\sigma)$ is dissipative because this coupling is proportional to $\gamma_\phi$.
Furthermore, other parts of viscous shear $\sim \partial^{<\mu} u^{\nu>}$ and volumetric expansion $(\partial \cdot u)$ influence the evolution of spin density. 
\vspace{0.5em}

\noindent
\textbf{Spin effects in steady states}
---
At global equilibrium, the entropy saturates its extremal value, i.e., $\partial_\mu(\mathcal{S}^\mu + \delta \mathcal{S}^\mu)=0$, then all the terms in the RHS of Eqs.~\eqref{eq:dissipative_02_1}, \eqref{eq:dissipative_02_2}, \eqref{eq:dissipative_02} vanish.  This requirement should recover the well-known Killing vector condition in relativistic ideal hydrodynamics~\cite{Gao:2012ix}, that is,
\begin{equation}
    \partial_{(\nu}(\beta u_{\mu)})=0 \,,
    \label{eq:KillingC}
\end{equation}
which is accompanied by $\partial\cdot u=0$, $\partial_{<\mu}u_{\nu>}=0$, $\Delta^{\mu\nu}\partial_{\nu}\beta + \beta \dot{u}^\mu=0$, and also spatially projected one of Eq.~\eqref{eq:GE} that constraints the spin potential and the thermal
vorticity tensor~\cite{Becattini:2012tc}. 
Let us discuss the $h^\mu -\mathcal{H} \nu^\mu, \pi^{\mu\nu}, \phi^{\mu\nu}$ in steady state described by the condition (\ref{eq:KillingC}).

Notably, the RHSs of Eqs.~\eqref{eq:dissipative_02_1}, \eqref{eq:dissipative_02_2}, \eqref{eq:dissipative_02} provide us with nonvanishing spin-induced terms of $-h_s^\mu$, $-\pi_s^{\mu\nu}$, and $-\phi_s^{\mu\nu}$, leading to the equilibrium values as 
\begin{align}
  h_{\text{eq}}^{\mu} - \mathcal{H}\nu_{\text{eq}}^{\mu}
  & = \frac{1}{2} \varepsilon^{\mu\nu\rho\sigma} \dot{u}_\nu u_\rho s_\sigma \notag\\
  &\qquad  - \frac{1}{2} \varepsilon^{\mu\nu\rho\sigma}\partial_\nu (u_\rho s_\sigma)  - u^\mu (s\cdot\vOmega)\,, \notag \\
  \pi_{\text{eq}}^{\mu\nu} & = \vOmega^{<\mu}s^{\nu>} + 2\Bigl(v_n^2 - \frac{1}{3}\Bigr) \Delta^{\mu\nu}(s\cdot\vOmega) \,, \notag \\
  \phi_{\text{eq}}^{\mu\nu} & = \vOmega^{[\mu} s^{\nu]}\,,
\label{eq:hydro_global}
\end{align}
where we have replaced $\omega^\mu$ with $\vOmega^\mu$ according to Eq.~\eqref{eq:GE}.

These are spin-induced corrections on top of conventional relativistic viscous hydrodynamics in which $h_{\text{eq}}^\mu-\mathcal{H}\nu_{\text{eq}}^\mu$ and $\pi_{\text{eq}}^{\mu\nu}$ would have vanished.  As we discuss below, nonvanishing $h_{\text{eq}}^\mu$ is interpreted as the inverse spin Hall effect and the anomalous Hall effect.  Let us take a closer look at $\phi_{\text{eq}}^{\mu\nu}$ here.

We see that the spin evolution equation~\eqref{eq:MOdifiedBMT} retains only the Thomas precession term at global equilibrium.  Actually, by imposing the equilibrium conditions~\eqref{eq:GE} and \eqref{eq:KillingC}, we find that the modified BMT equation~\eqref{eq:MOdifiedBMT} undergoes the following:
\begin{equation}
  \dot{s}^\mu_{\text{eq}}  =  -u^\mu s^{\nu}_{\text{eq}}\dot{u}_\nu \,.
  \label{eq:spin_global}
\end{equation}
Multiplying $g_{\mu\nu}s^\nu_{\text{eq}}$ to the both sides, 
we can show $\dot{s}^2_{\text{eq}}=0$, indicating that $s^{2}_{\textrm{eq}}$ is conserved at global equilibrium as expected.
Equation~\eqref{eq:spin_global} can provide us with additional constraints on $\phi^{\mu\nu}_{\text{eq}}$.  In global equilibrium states, the Lie derivative of any physical observable along $\beta u^\mu$ should vanish~\cite{Becattini:2016stj}, i.e., 
$\mathcal{L}_{\beta }s^{\lambda}_{\text{eq}} = \beta u^{\nu}\partial_{\nu}s^{\lambda}_{\text{eq}} - s^{\nu}_{\text{eq}} \partial_{\nu}(\beta u^\lambda)=0$,
which leads to 
$\dot{s}_{\text{eq}}^{\mu} = -u^{\mu}s_{\text{eq}}^{\nu}\dot{u}_{\nu} - \varepsilon^{\mu\nu\rho\sigma}{s_{\text{eq}}}_{\nu}u_{\rho}\vOmega_{\sigma}
$.
Comparing this with Eq.~\eqref{eq:spin_global}, we find that the second term should vanish, meaning that $\phi_{\text{eq}}^{\mu\nu} = \vOmega^{[\mu} s^{\nu]} = 0$.  Therefore, the spin density vector $\boldsymbol{s}$ is only parallel or antiparallel to the vorticity vector $\boldsymbol{\vOmega}$ in equilibrium.
\vspace{0.5em}

\noindent
\textbf{Non-relativistic limit and spin currents}
---
Let us discuss the physical interpretation of $h_{s}^{\mu}$.
In nonrelativistic limit, $u^{\mu}\approx(1,\,\bv), |\bv|\ll1$, we notice that $s^{\mu}\rightarrow(0,\,\bs)$ with $s^i=\frac{1}{2}\epsilon^{ijk}S^{jk}$.  Then, we get: 
\begin{equation}
  \begin{split}
  \boldsymbol{h}_{s} & = -\frac{1}{2}(\nabla\times\bs)  + \frac{1}{2}(\dot{\bs} \times \bv) + (1+\xi)(\bs \times  \dot{\bv}) \\
  &\qquad +\frac{\xi}{T}(\bs \times \nabla T) + \frac{\xi}{T}\dot{T}(\bs \times \bv) + \mathcal{O}(\bv^2)\,.
  \end{split}
 \label{eq:heat_flow_01-1}
\end{equation}
Remarkably, the first term $\propto \nabla\times\boldsymbol{s}$ in $\boldsymbol{h}_s$ recovers the inverse spin Hall effect~\cite{sinova2015spin}, and the other term $\propto \boldsymbol{s}\times\nabla T$ is interpreted as the anomalous Hall effect~\cite{nagaosa2010anomalous}.
Interestingly, besides these well-known spin effects, we also observe extra terms involving $\dot{\bm s}$ and $\dot{T}$ due to the relativistic nature.

These spin corrections have the potential to be further measured in tabletop experiments. 
In the energy flow frame (Landau frame), in which $u^\mu$ denotes the fluid velocity of the energy flow, we set $\boldsymbol{h}=0$ and observe the spin-induced correction to the current $\boldsymbol{\nu}$ as
\begin{equation}
  \boldsymbol{\nu}_s = \mathcal{H}^{-1}\boldsymbol{h}_{s}\,,
\end{equation}
as well as the standard heat conducting flow. 
In the particle frame (Eckart frame), in which $u^\mu$ is the fluid velocity of the particle motion, i.e., $\boldsymbol{\nu}=0$, the spin corrections should contribute to the energy flow, $\Theta^{0i}$, as
\begin{equation}
  \Theta^{0i}_s = -\boldsymbol{h}_{s}\,.
\end{equation}
\vspace{0.5em}

\paragraph*{Summary:}

We have derived a new form of relativistic spin hydrodynamics from the entropy principle, employing totally antisymmetric spin tensors.  We have found that the entropy production acquires additional contributions from the antisymmetric spin tensors, which can be decomposed into spin-induced corrections to the dissipative terms, such as heat flow and the viscous tensor, as well as to the antisymmetric part of the energy-momentum tensor.

We have formulated the evolution equation for the spin density, which should be interpreted as a generalization of the Bargmann–Michel–Telegdi (BMT) equation.  Importantly, the Thomas precession term emerges naturally from the antisymmetrized spin tensor, whereas the spin-rotation coupling required by thermodynamic consistency originates from the spin-induced correction to the antisymmetric part of the energy-momentum tensor.  These results ensure that well-established spin dynamics are consistently incorporated into our formulation of relativistic spin hydrodynamics.

Furthermore, we have taken the nonrelativistic equilibrium limit to confirm that the spin-induced contributions to heat-conducting flow are related to the inverse spin Hall effect and the anomalous Hall effect.  We have also identified new nondissipative terms arising from relativistic generalizations of spin effects, which lead to testable predictions.
Our findings have broad applicability in the study of relativistic many-body systems.
\vspace{0.5em}

\begin{acknowledgments}
The authors thank
Francesco~Becattini,
Wojciech~Florkowski,
Masaru~Hongo
for useful discussions.
K.F.\ thanks the Galileo Galilei Institute for Theoretical Physics for the hospitality and the INFN for partial support during the completion of this work.
K.F.\ was supported in part by JSPS KAKENHI under Grant No.\ 22H05118 and No.\ 23K22487.
S.P.\ was supported in part by National Key Research and Development Program of China under Contract No.\ 2022YFA1605500, Chinese Academy of Sciences (CAS) under Grant No.\ YSBR-088, and National Natural Science Foundation of China (NSFC) under Grant No.\ 12135011.
\end{acknowledgments}

\bibliography{Ref.bib}
\bibliographystyle{apsrev4-1}

\section*{End Matter}

We explain the concrete procedures to make the tensor decomposition for readers who are interested in deriving our expressions.  There are two unwanted terms in Eq.~\eqref{eq:divS}; namely, $2\beta\omega_{\mu\nu}S^{\lambda\mu}\p_\lambda u^\nu$ and $q^\mu(\p_\mu\beta - \beta\dot{u}_\mu)$.

Let us start with the first term, $2\beta\omega_{\mu\nu} S^{\lambda\mu}\p_\lambda u^\nu = -2\beta S^{\mu\lambda}\omega^\nu_{\;\lambda}\partial_\mu u_\nu$.
Using $\partial_\mu u_\nu = \partial_{(\mu} u_{\nu)} + \partial_{[\mu} u_{\nu]}$, we can decompose and put them together with others as
\begin{equation}
  \begin{split}
  & \beta(\pi^{\rho\sigma} -2 S^{(\rho\lambda} \omega^{\sigma)}_{\;\lambda})
  \Delta^\mu_\rho \Delta^\nu_\sigma \partial_{(\mu} u_{\nu)} \\
  & + (\phi^{\rho\sigma} -2 S^{[\rho\lambda} \omega^{\sigma]}_{\;\lambda})
  \Delta^\mu_\rho \Delta^\nu_\sigma (2\beta\omega_{\mu\nu}
  + \partial_{[\mu} \beta u_{\nu]})\,.
  \end{split}
\end{equation}
In this way, $-2\beta S^{\mu\lambda} \omega^\nu_{\;\lambda}\partial_\mu u^\nu$ can be absorbed by the renormalization in $\pi^{\mu\nu}$ and $\phi^{\mu\nu}$ as
$-2S^{(\mu\lambda} \omega^{\nu)}_{\;\lambda} \to \pi_s^{\mu\nu}$ and
$-2S^{[\mu\lambda} \omega^{\nu]}_{\;\lambda} \to \phi_s^{\mu\nu}$.

Now, let us turn to the term with $q^\mu$ in Eq.~\eqref{eq:divS}, which is more complicated.  With Eq.~\eqref{eq:qmu}, we can spell them out as
\begin{align}
  & \frac{1}{2}(S^{\mu\nu} \dot{u}_\nu + \Delta^\mu_\nu \partial_\lambda S^{\nu\lambda}) 
    (\partial_\mu \beta - \beta \dot{u}_\mu) \notag\\
  &= \frac{1}{2} S^{\mu\nu} \dot{u}_\nu \partial_\mu\beta
    + \frac{1}{2} \Delta^\mu_\nu \partial_\lambda S^{\nu\lambda} (\partial_\mu\beta - \beta\dot{u}_\mu )\,.
\end{align}
To extract the heat current renormalization, we should reexpress the above three terms into a form of $c(\partial_\mu\beta + \beta\dot{u}_\mu) + c'$.  However, it is easy to see that the decomposition of the first term cannot be determined uniquely from a trivial relation; $\bigl[ S^{\mu\nu}(\partial_\nu\beta+\beta\dot{u}_\nu) \bigr](\partial_\mu\beta + \beta\dot{u}_\mu) = 0$, implying that an arbitrary term $\propto S^{\mu\nu}(\partial_\nu\beta+\beta\dot{u}_\nu)$ is added to $h^\mu$ without changing entropy production.  This ambiguity in $h^\mu$ emerges from not necessarily the tensor decomposition.  The inclusion of antisymmetric tensor in the system inevitably leads to this ``nondissipative'' component in $h^\mu$.  In principle, we can remove this ambiguity from the physical requirement or the boundary condition.
For the moment, we rewrite it as $S^{\mu\nu}\dot{u}_\nu \partial_\mu\beta = S^{\mu\nu}\dot{u}_\nu (\partial_\mu\beta + \beta\dot{u}_\mu)$, so that $\frac{1}{2}S^{\mu\nu}\dot{u}_\nu \to h_s^\mu$.  In addition,
$\xi S^{\mu\nu}(\beta^{-1} \partial_\nu\beta + \dot{u}_\nu)\to h_s^\mu$ is allowed, where the ambiguity is expressed in an unfixed parameter $\xi$.

For the remaining two terms, there is again an ambiguity.  That is, $\Delta^\mu_\nu \partial_\lambda S^{\nu\lambda}(\partial_\mu \beta - \beta\dot{u}_\mu
) = \kappa\Delta^\mu_\nu\partial_\lambda S^{\nu\lambda}(\partial_\mu\beta + \beta\dot{u}_\mu) + (1-\kappa)\Delta^\mu_\nu \partial_\lambda S^{\nu\lambda}\partial_\mu\beta  -  (1+\kappa)\Delta^\mu_\nu \partial_\lambda S^{\nu\lambda} \beta\dot{u}_\mu$ with an arbitrary parameter $\kappa$.
The renormalization correction to the heat current from this contribution is
$\frac{\kappa}{2}\Delta^\mu_\nu\partial_\lambda S^{\nu\lambda} \to h_s^\mu$.
Here, we find that $\kappa=-1$ significantly simplifies the final results, though other $\kappa$ is still possible.
We can partially renormalize $\Delta^\mu_\nu \partial_\lambda S^{\nu\lambda} \partial_\mu \beta$ into the entropy current using
\begin{equation}
  \Delta^\mu_\nu \partial_\lambda S^{\nu\lambda} \partial_\mu \beta
  = \partial_\lambda (\Delta_\nu^\mu S^{\nu\lambda}\partial_\mu\beta) 
    - \dot{\beta} S^{\mu\nu} \partial_{\mu} u_{\nu}\,.
\end{equation}
In fact, because of $(u\cdot S)^\mu=0$, we see
$\Delta^\mu_\nu S^{\nu\lambda}\partial_\mu\beta = S^{\mu\lambda}\partial_\mu\beta$, which means 
$\delta\mathcal{S}^\mu = S^{\mu\nu} \partial_\nu\beta$.
Finally, the last remaining term is decomposed as
\begin{equation}
  - \dot{\beta} S^{\mu\nu} \partial_{\mu} u_{\nu}
  = -\dot{\beta} S^{\mu\nu} (2\omega_{\mu\nu} + \partial_\mu u_\nu)
    + 2\dot{\beta} S^{\mu\nu}\omega_{\mu\nu}\,.
\end{equation}
Here, the first term gives additional renormalization to $\phi^{\mu\nu}$ as
$-\beta^{-1}\dot{\beta} S^{\mu\nu}\to \phi_s^{\mu\nu}$.
We can put the second term together with the bulk viscous term using
\begin{equation}
  \dot{\beta} = v_n^2 \beta (\partial\cdot u) + \mathcal{O}(\partial^2)\,,
\end{equation}
where $v_n^2=(\partial p/\partial e)_{n,S^{\mu\nu}}$ as defined in the main text.
Now, we have completed our demonstration to assure that $\partial_\mu \mathcal{S}^\mu\ge 0$ can be guaranteed even when we constrain $q^\mu$ in Eq.~\eqref{eq:qmu}.

\end{document}